# Network learning via multi-agent inverse transportation problems

Susan Jia Xu[1], Mehdi Nourinejad[2], Xuebo Lai[3], Joseph Y. J. Chow[1*]

[1]Department of Civil & Urban Engineering, New York University, New York, NY, USA
[2]Department of Civil & Mining Engineering, University of Toronto, Toronto, ON, Canada
[3] Department of Computer Science, Courant Institute, New York University, New York, NY, USA

[*]Corresponding author email: joseph.chow@nyu.edu

## Abstract

Despite the ubiquity of transportation data, methods to infer the state parameters of a network either ignore sensitivity of route decisions, require route enumeration for parameterizing descriptive models of route selection, or require complex bilevel models of route assignment behavior. These limitations prevent modelers from fully exploiting ubiquitous data in monitoring transportation networks. Inverse optimization methods that capture network route choice behavior can address this gap, but they are designed to take observations of the same model to learn the parameters of that model, which is statistically inefficient (e.g. requires estimating population route and link flows). New inverse optimization models and supporting algorithms are proposed to learn the parameters of heterogeneous travelers' route behavior to infer shared network state parameters (e.g. link capacity dual prices). The inferred values are consistent with observations of each agent's optimization behavior. We prove that the method can obtain unique dual prices for a network shared by these agents in polynomial time. Four experiments are conducted. The first one, conducted on a 4-node network, verifies the methodology to obtain heterogeneous link cost parameters even when multinomial or mixed logit models would not be meaningfully estimated. The second is a parameter recovery test on the Nguyen-Dupuis network that shows that unique latent link capacity dual prices can be inferred using the proposed method. The third test on the same network demonstrates how a monitoring system in an online learning environment can be designed using this method. The last test demonstrates this learning on real data obtained from a freeway network in Queens, New York, using only real-time Google Maps queries.

# 1 Introduction

Travel data has become increasingly abundant and ubiquitous in recent years due to advances in information and communications technologies (ICTs) and Big Data. The shift to data-driven methods in transportation is apparent. Notable efforts include the Mobile Millennium project to track traffic using GPS phone data (Herrera et al., 2010), use of transit smart card data in Santiago to estimate travel demand patterns (Munizaga and Palma, 2012), urban link performance inference from New York City taxi data (Zhan et al., 2013), and the Digital Matatu Project in Nairobi, Kenya, to characterize their flexible bus system with cell phone data (Williams et al., 2015), among others.

Such methods make use of observations from data to infer or learn specific characteristics of system components that are not directly observable. For example, the study from Zhan et al. (2013) infers link travel times using taxi GPS data. The methodology is generally classified as an inverse model (see Tarantola, 2005), which assumes there exists a model $M$ that transforms a set of parameters $\theta$ to a set of outputs $X$ as $X = M(\theta)$. The inverse model deals with finding a set of parameter estimates $\hat{\theta}$ based on observed outputs $x$ as $\hat{\theta} = M^{-1}(x)$. In a network setting, the state of a network may be defined by several types of parameters that require estimation: link costs, link/path flows, origin-destination (OD) flows, link capacities or congestion effects.

Inference in a network context using inverse models has a long history and literature. The primary factor distinguishing inference methods for network parameters is whether, and how, the method handles sensitivity of route decisions by travelers or (in the case of non-transportation networks) packets to the state of the network. A second consideration is that system parameters (e.g., link costs and capacity effects) need to be inferred as they depend on external factors like weather, presence of incidents, etc., particularly in transportation networks.

The simplest inference methods relate an observable set of attributes to the desired parameters directly through network structure (e.g. link flow is the sum of the proportions of all OD flows that use that link) absent of any route decision sensitivity. One example is Van Zuylen and Willumsen (1980), who propose an OD estimation model from link traffic counts using entropy maximization to determine the proportion of OD flows that are most likely to traverse an observed link. The variables are connected by an incidence matrix that reflects the network structure. Instead of using entropy maximization, Bell (1991) uses a constrained generalized least squares approach with additional information from prior surveys/studies to regularize the inverse problem. Path flows are not explicit in such models.

Inverse models that deal with route sensitivity do so by explicitly mapping observed variables like link counts to latent path flows, and then mapping those path flows to desired parameters like OD flows. Examples include Vardi (1996) and Tebaldi and West (1998). Vardi (1996) coins this topic as "network tomography", and proposes methods to estimate parametric distributions of flow from observed link count data for a single known path or for a Markovian distribution of paths. Tebaldi and West (1998) use Bayesian inference to estimate the parametric distributions considered by Vardi (1996). Since these methods require path enumeration, it may be difficult to apply them to large networks. Recent efforts (Airoldi and Blocker, 2013; Hazelton, 2015) use more efficient route flow sampling strategies to make these inference models more scalable. These methods are tested on networks with up to 21 nodes.

In the case of transportation networks, two qualities offer unique challenges and opportunities to network inference: (1) state changes can involve system parameters like link capacities due to weather or incidents; and (2) route decisions are fundamentally governed by *behavioral mechanisms* like route choice behavior. Changes that involve link capacities or other similar system parameters may be much harder to anticipate using parametric models (e.g. normal distributions) for path assignment based only on prior data. To compensate for this, transportation researchers have inserted behavioral mechanisms (Cascetta et al., 1996; Vovsha and Bekhor, 1998; Srinivasan and Mahmassani, 2000; Dia, 2002; Frejinger and Bierlaire, 2007; Ben-Elia and Shiftan, 2010; Gao, 2012; Fosgerau et al., 2013) into the inverse models so that route sensitivity can rely on the mechanism where data is sparse. Route enumeration remains an issue in some cases, even as researchers seek ways to address that with choice set selection. For example, Fosgerau et al. (2013), Baillong and Cominetti (2006), and Akamatsu (1996) make use of Markovian models to overcome route generation. A literature survey is available from Prato (2009).

To overcome the route enumeration problem, more integration of behavior with network structure has been sought (see Watling et al., 2015). For example, when considering congestion or capacity effects of networks, the presence of a stochastic user equilibrium behavior (Daganzo and Sheffi, 1977) may be assumed. In the case of OD estimation, Yang et al. (1992) assume Wardrop's user equilibrium behavioral principle to assign observed link flows to route flows without having to enumerate paths. Other efforts include Ashok and Ben-Akiva (2002), who allow for perturbations from the behavioral mechanism with stochastic assignment. A new problem emerges, however, with the complexity of having a nonconvex bilevel optimization problem for the inverse model.

To summarize, methods to infer the state parameters of a network either ignore sensitivity of route decisions, require route enumeration for parameterizing models of route selection, or require complex bilevel models of route assignment behavior. Some of these inference methods, particularly for demand patterns in a network, may not even be necessary in an age of data ubiquity. For instance, methods like Tang et al. (2015) and Alexander et al. (2015) use GPS or mobile phone data to reduce OD estimation to a basic sampling problem without having to infer them from other observed variables. Nevertheless, there remains a challenge of being able to explain network state changes, as illustrated in Fig. 1. The classic Nguyen-Dupuis network (Nguyen and Dupuis, 1984) is used to show two states of latent link capacities imposed on the network. In the figure, the change from one state on the left to the next on the right is due to a change in the link 7 capacity. Based on observation (and hypothetically from methods in Tang et al. (2015) and Alexander et al. (2015)), we can compute or estimate the total system travel times (68,400 on left, 70,000 on right) and observe that there is a difference in flow on links 3, 4, 5, 7, 8, 9, 10, 11, 12, 14, and 15, but we cannot explain *why* the state change occurs.

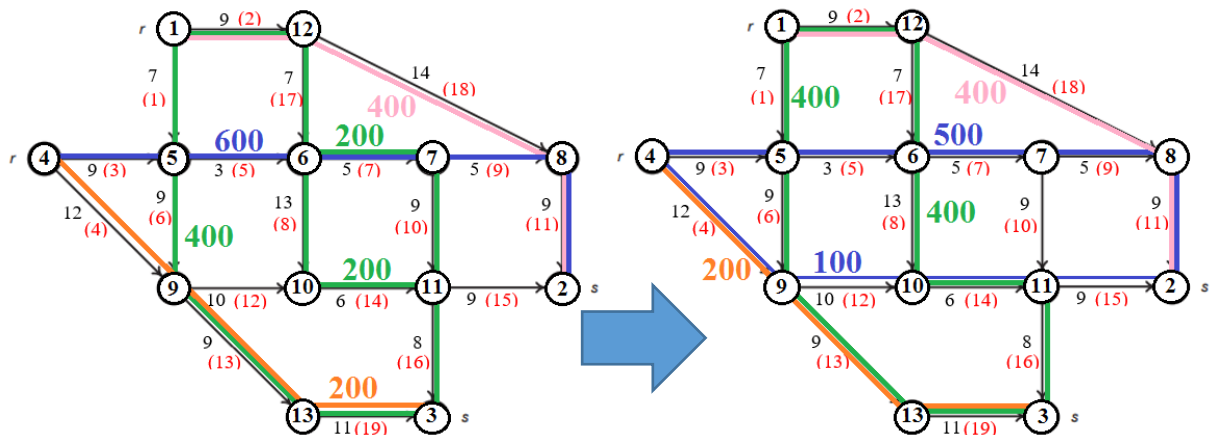

**Fig. 1. Illustration of a state change in the Nguyen-Dupuis network that leads to a change in observed flows (link IDs in red parentheses).**

One potential solution for network inference is *inverse optimization*. Ahuja and Orlin (2001) proposed a class of inverse problems posed as a linear optimization model. This class of problems, called inverse optimization (IO), allows one to learn linear programming parameters from observed decision variables and prior information such that the desired optimality condition is maintained. IO models capture the route choice mechanism if it follows an assignment-based mathematical program. The models also have great flexibility in inferring either demand or system parameters.

Nonetheless, inverse optimization in the current literature also cannot effectively explain the state change in Fig. 1 for two primary reasons. First, the closest model from Güler and Hamacher (2010) studies the single commodity capacity inverse minimum cost flow problem and conclude that it is NP-hard. However, the example in Fig. 1 describes a multicommodity flow problem in which there is demand from node 1 and node 4 to node 2 and node 3. Second, IO assumes that the data obtained is at the same level of the model, e.g. an inverse of a network flow model uses observation of the flow data to learn the parameters of the system model. In many transportation cases, however, this is untrue. The system model is the whole transportation network, and its parameters dictate how the congestion and capacity effects influence travel. However, it is the individual agents that are optimizing their own network models (e.g. shortest path) as a behavioral mechanism and learning from their experiences. Without addressing this discrepancy, current IO methods need to estimate variables at the population level, e.g. total link or path flows, which is statistically inefficient. If we are able to synchronize the route choice behavioral mechanism at the system level with the behavior of individual agents, we may be able to avoid this costly step.

We propose to address this discrepancy with a new data-driven methodology that uses inverse optimization with network models that rely on only learning from sampled heterogeneous agents. The method takes observed routes of travelers to estimate heterogeneous route preferences and infer network parameters that influence these observed routes, such as the dual prices of the link capacities. Because the method learns the preferences of multiple agents and relates those preferences to system parameters, it can be used in an online learning environment in which system parameters are updated over time based on *only* real-time sampling of individual agents without having to estimate population link or path flows.

The remainder of the paper is organized as follows. Section 2 expands on the review of inverse optimization as a learning methodology, and recent developments to handle heterogeneity in observed data. Section 3 presents the proposed methodology in two stages: first as a basic model to infer heterogeneous route preferences of multiple agents using a shared network, and second in inferring the parameters of that network to explain their route choices. Section 4 presents four computational experiments that include a demonstration of online learning in a synthetic network and in a real data illustration of a freeway network in Queens, NY. Section 5 concludes.

## 2 Review of inverse optimization

Inverse optimization can be used to learn network parameters from a prior value. The first inverse optimization model was proposed by Burton and Toint (1992) for the inverse shortest path problem, and it was further generalized by Ahuja and Orlin (2001). It is defined as follows: for a given prior $c_0$ of a linear program's (LP's) parameters and observed decision variables $x^*$, determine an updated $c$ such that $x^*$ is optimal while minimizing its $L_1$ norm from the prior, as shown in Eq. (1). $A$ is the constraint matrix and $b$ is the vector of side constraint values.

$$\min_c |c_0 - c| : x^* = argmin\{c^T x : Ax \leq b, x \geq 0\} \tag{1}$$

The $L_1$ norm minimization is used to regularize what would otherwise be an ill-posed problem with infinite solutions. This ensures that a unique solution can be obtained given a prior $c_0$. Alternatively, weights can be added for further calibration. Ahuja and Orlin (2001) showed that Eq. (1) (as well as the $L_\infty$ norm variant) can be reformulated as an LP. This is done by introducing two non-negative decision variable vectors $e$ and $f$ such that their difference is equal to $c_0 - c$: $c_0 - c = e - f$. The problem can then be reformulated using strong duality and dual feasibility conditions, as shown in Eq. (2) to Eq. (5).

$$\min_{y,e,f} e + f \quad (L_1 \text{ norm minimization}) \tag{2}$$

subject to

$$A^T y \geq c_0 - e + f \quad (dual\ feasibility) \tag{3}$$
$$b^T y = (c_0 - e + f)^T x^* \quad (strong\ duality) \tag{4}$$
$$y, e, f \geq 0 \quad (non\text{-}negativity) \tag{5}$$

where $y$ is a dual variable of the original LP.

There have been a number of advances and applications in inverse optimization, with an early survey on combinatorial IO by Heuberger (2004). Table 1 provides a summary of these advances. For example, Wang (2009) proposed a cutting plane method to solve the inverse mixed integer linear programming (InvMILP) problem. Güler and Hamacher (2010) proposed an IO model for estimating link capacities. Zhang and Zhang (2010) proposed an inverse quadratic program. Chow and Recker (2012) proposed an inverse vehicle routing problem (VRP) that includes side constraint (goal arrival times) estimation. Bertsimas et al. (2015) proposed an inverse variational inequality to estimate the parameters that would lead to observed patterns under equilibrium.

There are also many applications for IO. Day et al. (2002) used IO as a tool for model calibration for railroad networks. Burkard et al. (2004) applied IO to facility location problems. Agarwal and Ergun (2008) used IO as a mechanism design approach to select a solution from the core for a multicommodity flow game between different service networks seeking an alliance. Brucker and Shakhlevich (2009) studied applications in inverse scheduling. Bertsimas et al. (2012) applied IO to estimate the Black-Litterman model in portfolio optimization in the financial industry. Birge et al. (2014) applied IO to reveal the electricity market structure based on observed choices made by competitors. Chow et al. (2014) proposed an inverse traffic assignment problem for calibrating impedances at freight facilities, and You et al. (2016) used truck GPS data to infer parameters to forecast urban truck delivery patterns. Hong et al. (2017) proposed using inverse optimization to infer the heterogeneous route-level parameters of a mixed logit model of route choice, and tested that method with transit smart card data taken from 50,000 trips in the Seoul metro system. They show empirically that the fixed point method proposed in Chow and Recker (2012) is able to discern heterogeneous parameter distributions for a utility function composed of transit time, transfer, and crowding.

**Table 1. Overview of inverse optimization advances and applications**

| Methodological advances | | New applications | |
|---|---|---|---|
| Burton and Toint (1992) | Inverse shortest path | Day et al. (2002) | Network calibration |
| Ahuja and Orlin (2001) | Inverse linear programming | Burkard et al. (2004) | Inverse median problem |
| Wang (2009) | Inverse MILP | Agarwal and Ergun (2008) | Mechanism design |
| Güler and Hamacher (2010) | Link capacities in inverse minimum cost flow problem | Brucker and Shakhlevich (2009) | Inverse scheduling |
| Zhang and Zhang (2010) | Nonlinear inverse optimization | Bertsimas et al. (2012) | Financial portfolio management |
| Chow and Recker (2012) | Multi-agent inverse optimization, inverse VRP with side constraint estimation | Birge et al. (2014) | Electricity market structure |
| Aswani et al. (2015) | Noisy data | Chow et al. (2014) | Inverse traffic assignment |
| Bertsimas et al. (2015) | Inverse variational inequality | You et al. (2016) | Urban truck forecasting |
| Esfahani et al. (2015) | Incomplete information | Hong et al. (2017) | Mixed logit estimation |

Despite these developments, IO methods involve a single model and observations of decision variable outputs of that model. In many instances, observations are not made at the aggregate level, but come from individual behavioral agent decisions. This dichotomy between a system model and learning through agent observations is an important distinction. First, IO methods require system observations, but obtaining that information from individual agents result in heterogeneous inputs. The result is that additional estimation of system aggregation from agent observations is needed, which leads to increased inefficiencies. Second, the resulting information from agent observations may not be consistent with the system model when inferring system parameters. This limitation is evident in the IO literature, where most efforts have focused on systems with learning from the same system level observations (e.g. observed company choices and their electricity pricing decisions, or an individual's perceived costs of a network from their own observed path choices).

Recent studies have tried to address these point by using noisy observations. Aswani et al. (2015) set up a bilevel problem to estimate from noisy data, and Esfahani et al. (2015) modeled the noisy information problem as a robust optimization model. Both approaches explain the heterogeneity with stochastic distributions to allow suboptimal observations. However, this leads to discrepancies with mechanistic assumptions in the prevailing system model (i.e. it may lead to an observation that is not optimal with respect to their own preferences). The limitation of current IO methods to using system observations is also in computational efficiency. For example, Güler and Hamacher (2010) proposed an IO model to infer link capacities for a minimum cost flow problem from observed flows. They conclude that the model is NP-hard.

Chow and Recker (2012) proposed a multi-agent framework for IO where a sample of individuals' trip scheduling data is obtained and used to infer parameters of individual activity scheduling (variant of vehicle routing from Recker, 1995). In this case, parameters of multiple individuals are estimated such that the mean of their parameters is a fixed point. This leads to a learning process for heterogeneous parameters of a system model, where individually calibrated optimization models correspond to observations as optimal solutions. Chow and Djavadian (2015) showed how sampled heterogeneous parameters can be fit into a mixed logit representation of constrained activity schedule choice. Hong et al. (2017) empirically proved the effectiveness of using inverse optimization for mixed logit route choice model parameter estimation with transit smart card data.

We formalize the multi-agent approach for network route choice such that observations are taken from agents, estimated parameters are heterogeneous and consistent with those agent observations, and further used to infer system-level network parameters that impact those observations.

## 3 Proposed methodology

We first formalize the multi-agent IO framework for general networks with route assignment, and then expand that framework to a methodology to infer network parameters.

### 3.1 Basic multi-agent inverse transportation problem framework

Consider a network $G(N,A)$ that receives observations from a population $P$ of agents behaviorally seeking to travel from an origin $r_i \in N$ to a destination $s_i \in N$, $\forall i \in P$ according to a shortest path in terms of additive link costs. Each agent $i \in P$ has a perception of network parameters in a subnetwork $g_i \subseteq G$; these varying perceptions are reflected in heterogeneous parameters at the system level.

In the basic multi-agent inverse transportation problem framework, let us assume there are no congestion or capacity effects, and only heterogeneous link costs are present. In other words, each link cost $c_a, a \in A$, is described by a distribution over $P$ such that an agent's perceived values of $c_{a,i}$ justify their revealed route choice $x_i^*$. Parameter learning is achieved with a set of inverse shortest path problems $\phi^{-1}(g_i, c_0, x_i^*)$, one for each agent, constrained to have an invariant common prior, as illustrated in Fig. 2 and in Eq. (6) and Eq. (7).

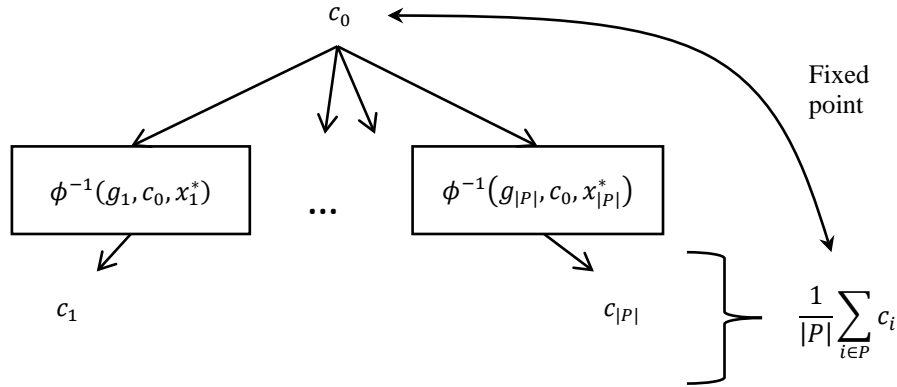

**Fig. 2. Illustration of multi-agent inverse optimization as a fixed point problem.**

$$\min_{c_0, c_i} \{|c_0 - c_i|: x_i^* = argmin_x \, \phi(g_i, c_i)\}, \quad \forall i \in P \tag{6}$$

Subject to

$$c_0 = \frac{1}{|P|} \sum_{i \in P} c_i \tag{7}$$

Since $\phi$ is a shortest path problem, which is an LP, each agent's inverse LP (InvLP) can reach a unique LP solution under $L_1$ norm (Ahuja and Orlin, 2001) for a given initial prior $c_0$. The constraint in Eq. (7) interprets the prior as a common prior (Feinberg, 2000; Chow and Recker, 2012) among the population. Given an initial guess of a common prior, a convergent iterative algorithm (e.g. Method of Successive Averages (MSA)) would reach a unique and statistically consistent fixed point with respect to that guess (see Chow and Recker, 2012). If we change the starting point (for example shifting all $c_0$ values from all 1 to all 1000), it would lead to a different fixed point solution set. This is similar to how the estimated parameters in random utility models are also only unique relative to each other, but the overall values can be scaled up or down. An MSA-based algorithm is shown here.

**Algorithm 1: MSA-based algorithm to solve Eq. (6) to Eq. (7)**
0. Given an initial common prior $c_0^1$ (e.g. previous update), and set $n = 1$.
1. For each agent $i \in P$, solve an inverse shortest path problem $c_i^n = \phi^{-1}(g_i, c_0^n, x_i^*)$.
2. Set average to $\mu^n = \frac{1}{|P|}\sum_{i \in P} c_i^n$.
3. Update common prior: $c_0^{n+1} = \frac{n}{n+1}c_0^n + \frac{1}{n+1}\mu^n$. Set $n = n + 1$ and go to step 1 if stopping criterion not reached.

If a sample $S \subset P$ of the population is used to infer the fixed point and distribution of the heterogeneous parameters, Chow and Djavadian (2015) show that a constrained mixed multinomial logit utility function as shown in Eq. (8) can aggregate agent observations. In a discrete choice model of route choice, each individual selects a route that maximizes their utility. The utility function can correspond to the objective function estimated for the individuals' shortest path problems. Consider Eq. (8) as a random utility function representing a whole population.

$$U_{ji} = \beta_j X_{ji} + \varepsilon_{ji} \qquad (8)$$

where $U_{ji}$ is the utility for alternative (route) $j$ by agent $i$, $\beta_j$ is a random coefficient vector that is normally distributed, $X_{ji}$ is an attribute of the alternative, and $\varepsilon_{ji}$ is a Gumbel-distributed error. To compute the probability from the normal distribution of the random coefficients, one can simulate a set of $R$ draws such that a simulated probability can be computed: $\Pr(j|\beta_j, X_{ji}) = \frac{1}{|S|}\sum_{r \in R} \Pr(j|\beta_{jr}, X_{ji})$. For the $S$ agents, the objective function of each agent is equivalent to a "random draw" of the random coefficients $c_i$ of the utility function in the mixed logit model if we assume $S = R$.

While the discussion here focuses on link travel costs only, the framework is applicable to other types of network flow parameters. For example, in Chow and Djavadian (2015), activity routing models capture parking duration, schedule delay, schedule makespan, and number of trip chains in the objective function. Hong et al. (2017) estimate the transit time, transfer, and crowding parameters in their route choice model.

*3.2 Network parameter inference via decomposition*
The methodology is further expanded to directly consider network parameters. Consider the capacitated multicommodity flow problem in Eq. (9) to Eq. (12), where $M$ is the set of commodities and $u = \{u_{a \in A}\}$ is a vector of capacity constraints for a subset of links in the network.

$$\min_{x} \sum_{m} c^T x_m \quad \text{(cost minimization)} \quad (9)$$

subject to

$$A x_m = b_m, \quad \forall m \in M \quad \text{(flow conservation for each commodity } m\text{)} \quad (10)$$

$$\sum_{m \in M} x_m \leq u \quad \text{(bundled capacity constraints)} \quad (11)$$

$$x_m \geq 0, \quad \forall m \in M \quad \text{(non-negativity)} \quad (12)$$

The inverse problem to infer the values of $u$ from observed $x_m$ and other network parameters is NP-hard (Güler and Hamacher, 2010). Instead of tackling this inverse problem directly, we seek dual prices $w$ corresponding to the constraint set (11). A value of $w_a = 0$ means that a link $a \in A$ is not operating at capacity $u_a$, while $w_a > 0$ reflects the impact of a binding capacity on agents' route choices. In other words, we do not concern ourselves with finding capacity, but instead with finding the effects of the capacity and its interaction with the agents.

In this problem, we assume link costs are not heterogeneous and known in advance. Instead, each agent has a perceived value of the dual price of the capacitated links. The capacitated problem can be decomposed into a master problem for determining optimal dual prices and unconstrained sub-problems for each commodity. The dual price is reflected within each agent's shortest path problem through the Partial Dualization Theorem (Ahuja et al., 1993): the $w$ corresponding to Eq. (11) in the multicommodity flow problem is equivalent to a $w$ for the uncapacitated shortest path problem of each agent $i \in P$ as shown in Eq. (13).

$$\min_{x_i} (c + w)^T x_i \quad (13)$$

By relying on this relationship, we introduce a multi-agent inverse transportation problem to infer the network dual prices. Each agent solves an IO where there is a common prior dual price vector $w_0$. We define two non-negative decision variables $e_i$ and $f_i$ for each agent such that $w_0 - w_i = e_i - f_i$, and solve Eq. (14) to Eq. (18) for each agent, subject to Eq. (19) for all agents. In other words, route dependencies are captured by bundle constraints such as capacity (Eq. 11). With decomposition, the original problem is decomposed into individual shortest path problems where the costs in the objective are updated to reflect the dual price obtained from the restricted master problem (Eq. 13). In the inverse problem, the requirement for a common prior (Eq. 19) ensures the solution will fit the bundling constraints.

$$\min_{y_i, e_i, f_i} e_i + f_i, \quad \forall i \in P \quad \text{($L_1$ norm minimization for each agent } i \in P\text{)} \quad (14)$$

subject to

$$A^T y_i \geq c + w_0 - e_i + f_i, \quad \forall i \in P \quad \text{(dual feasibility)} \quad (15)$$

$$b^T y_i = (c + w_0 - e_i + f_i)^T x_i^*, \quad \forall i \in P \quad \text{(strong duality)} \quad (16)$$

$$w_0 - e_i + f_i \geq 0, \quad \forall i \in P \quad \text{(dual price non-negativity)} \quad (17)$$

$$y_i, e_i, f_i \geq 0, \quad \forall i \in P \quad \text{(non-negativity)} \quad (18)$$

$$w_0 = \frac{1}{|P|} \sum_{i \in P} w_i \quad \text{(common prior)} \quad (19)$$

The formulation in Eq. (14) to Eq. (18) refers to an inverse of a generic LP in standard form as expressed in Eq. (1). In the case of taking the inverse shortest path, there are equality constraints so the dual variables for that problem are unbounded. The following three assertions are made.

**Proposition 1**. *Eq. (14) to Eq. (19) has a unique solution in a common prior dual price vector for all capacitated links, and this vector is the same for all agents, i.e. $w_0 = w_i \; \forall i \in P$.*

**Proof.** A multicommodity flow problem solution has a unique set of dual prices (Ahuja et al., 1993). This homogeneity occurs because the dual price is a lower bound threshold for each individual, and the highest value price is the one kept. This can be illustrated with two agents A and B sharing a link $a$. Suppose agent A would leave link $a$ if the dual price was $w_A$. This means any value of $w \geq w_A$ would incentivize agent A to leave link $a$. Now suppose agent B has a dual price of $w_B > w_A$. Any common prior price $w_A \leq w_0 < w_B$ would not be fixed, because agent B would perturb up towards $w_B$ while agent A would be indifferent, until the common prior and final prices become fixed at $w_0 = w_B$, and both agent A and B share the same $w_B$. ∎

**Proposition 2.** *The unique inverse optimal parameters to Eq. (14) to Eq. (19) can be reached by starting with an initial guess at $w_0^1 = 0$ and then following a basic iterative update of $w_0^{n+1} := \frac{1}{|P|} \sum_{i \in P} w_i^n$.*

**Proof**. Since $w_0^1 = 0$ represents the lower boundary, in each iteration $n$ the updated average of $w_0^n$ would always be increasing due to the lower threshold condition explained in the Proposition 1 proof. This means a basic iterative update of letting $w_0^{n+1} := \frac{1}{|P|} \sum_{i \in P} w_i^n$ is monotonically increasing. Therefore, it is guaranteed to reach the unique solution. ∎

The algorithm is explicitly shown here.

**Algorithm 2: Iterative algorithm to solve Eq. (14) to Eq. (19)**
0. Given an initial common prior $w_0^1$ (e.g. previous update), and $n = 1$.
1. For each agent $i \in P$, solve an inverse shortest path problem with augmented link costs in Eq. (13), $w_i^n = \phi^{-1}(g_i, w_0^n, x_i^*)$.
2. Update common prior: $w_0^{n+1} = \frac{1}{|P|} \sum_{i \in P} w_i^n$. Set $n = n + 1$ and go to step 1 if $w_0^{n+1} \neq w_0^n$.

**Proposition 3.** *The unique inverse optimal parameters to Eq. (14) to Eq. (19) can be reached in polynomial time using the basic iterative update from Proposition 2.*

**Proof**. Each run of the agent IO problem is an LP which is polynomial time solvable. The number of iterations of the iterative update is finite. This can be shown in a worst case scenario; suppose out of $|P|$ agents, $|P|-1$ of them all exhibit dual price of 0 for a particular link while one agent $i$ has a dual price of $w_i > 0$. In this case, in each iteration all the $|P|-1$ agents would keep setting the $w$'s to 0 and agent $i$'s to $w_i$. This means in the worst case the average will always be increasing by $\frac{w_i}{|P|}$ as a finite step size until the optimum is reached. ∎

These properties of the methodology signify the effectiveness of using agent observations to learn network parameters. We illustrate the methodology for two iterations. Consider three link flows observed in the network in Fig. 3, $x = \{100, 200, 100\}$. We can assume there are three groups of homogeneous agents, agent group 1 choosing link 1, agent group 2 choosing link 2, and agent group 3 choosing link 3. Each agent group seeks a dual price to explain their link choice, resulting in nine values of $w$ (for each agent and each link), and three values of $w_0$.

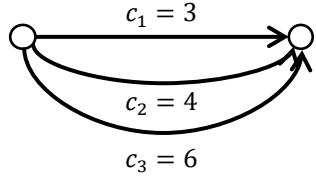

**Fig. 3. Toy network used for illustrating methodology.**

The algorithm is initiated by setting $w_0^1 = \{0,0,0\}$. An inverse shortest path problem is run for each agent. For agent group 1, $w_1^1 = \{0,0,0\}$ because they are already traveling on the shortest path with dual prices at zero. For agent group 2 to choose link 2, a value of $w_2^1 = \{1,0,0\}$ is needed. Lastly for agent group 3 to choose link 3, a value of $w_3^1 = \{3,2,0\}$ is needed. At the end of this iteration, the weighted average of the three agents is taken as the new prior: $w_0^2 = \left\{\frac{200+300}{400}, \frac{200}{400}, 0\right\} = \left\{\frac{5}{4}, \frac{1}{2}, 0\right\}$. If this is advanced a second iteration, we would get $w_1^2 = \left\{\frac{5}{4}, \frac{1}{2}, 0\right\}$, $w_2^2 = \left\{\frac{3}{2}, \frac{1}{2}, 0\right\}$, and $w_3^2 = \{3,2,0\}$. These would lead to a new prior $w_0^3 = \left\{\frac{(125+300+300)}{400}, \frac{50+100+200}{400}, 0\right\} = \left\{\frac{29}{16}, \frac{7}{8}, 0\right\}$. By inspection we can see that the dual prices will approach $w_0^* = w_1^* = w_2^* = w_3^* = \{3,2,0\}$.

### 3.3 Online learning

Suppose we have a system that receives agent routes as they are revealed in real time. The multi-agent IO model works in that setting without having to estimate population level parameters. It is assumed that each time a traveler updates their route decision, that information is sent to a system that learns from the observation to update link capacity dual prices. This information can then be used to monitor how system changes are affecting traveler decisions in real time. We augment Algorithm 2 to this setting as Algorithm 3.

**Algorithm 3: online learning algorithm to update system**
0. Given: an initial common prior (obtained from a system) $w_0^i$.
1. For newly arrived agent $i \in P$, solve an inverse shortest path problem with augmented link costs in Eq. (13), $w_i^{i+1} = \phi^{-1}(g_i, w_0^i, x_i^*)$.
2. Update common prior: $w_0^{i+1} = w_i^{i+1}$.

# 4 Numerical experiments

Four experiments are conducted. The first is performed on a small network to evaluate the proposed method without capacity effects. The second and third tests are conducted on the Nguyen-Dupuis network with capacity effects. We perform a parameter recovery test to see whether hidden dual prices can be recovered using the methodology. We also verify that the method can be applied in an online multi-agent learning setting. In the fourth test, the online learning is demonstrated using real data from a freeway network in Queens, New York City, and Google Maps real-time shortest path queries over a 3-hour period. All the data sets generated for these tests are publicly accessible on https://github.com/BUILTNYU/Network-learning-via-multi-agent-inverse-transportation-problems.

## 4.1 Verification of method to estimate heterogeneous link cost parameters
### 4.1.1 Experiment 1 design

This experiment has two primary objectives. The first is to illustrate the capability of the proposed method to capture heterogeneity of users' preferences at one network level (*link* costs) even when observations are made at another level (*route* choice). This objective is achieved by using a simple network with enumerated paths, and simulated link costs that vary across the population. These link cost variations reflect different traffic and environmental conditions (e.g. weather and road surface conditions) present during each user's trip, while observable route choices may be obtained from GPS, phone, or transit smart card data (to varying degrees). Link cost heterogeneity is reflected in distributions of the link costs across the population.

The second objective is to demonstrate how the proposed method can better handle structural changes in the underlying network. This is accomplished by applying the estimated models on a scenario where one of the links is removed.

To give the results more context, we estimate two discrete choice models: an aggregate multinomial logit model for route choice, and a mixed multinomial logit model that allows distributions in the path cost taste parameter. In total, the following scenarios shown in Table 2 are evaluated. Parameter estimation is run for the first six scenarios.

Table 2. Scenarios evaluated in Experiment 1

| No. | Scenario | Model |
|---|---|---|
| 1 | Baseline, independent links | Multinomial logit |
| 2 | | Mixed multinomial logit |
| 3 | | Shortest path problems calibrated with Algorithm 1 |
| 4 | Baseline, correlated links | Multinomial logit |
| 5 | | Mixed multinomial logit |
| 6 | | Shortest path problems calibrated with Algorithm 1 |
| 7 | Link 3 removed, independent links | Multinomial logit |
| 8 | | Shortest path problems calibrated with Algorithm 1 |
| 9 | Link 3 removed, correlated links | Multinomial logit |
| 10 | | Shortest path problems calibrated with Algorithm 1 |

### 4.1.2 Experiment 1 data

Consider a network as shown in Fig. 4 with five links identified in blue, where there are 500 agents traveling from node 1 to node 4. For the independent links scenario, the perceived link costs of the 500 agents are randomly simulated resulting in mean link costs of $\bar{c} = (0.49, 0.50, 0.50, 0.48, 0.49)$ and standard deviations of $\sigma_c = (0.29, 0.28, 0.29, 0.29, 0.28)$. The data is available on the GitHub site noted in Section 4 as *Test Set 1*. There are only three paths in their choice set represented by the following link sequences: (1,4), (2,5), (1,3,5), where their average path costs are 0.97, 0.99, and 1.48, respectively. Based on the simulated perceived costs and assumption that the travelers choose shortest routes, 48% choose (1,4), 48% choose (2,5), and 4% choose (1,3,5).

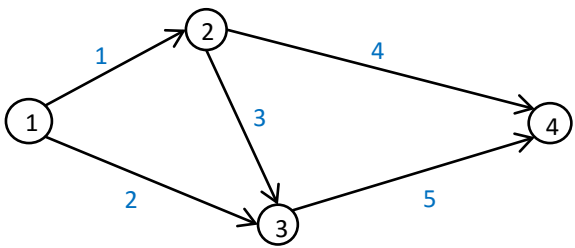

Fig. 4. Test network for Section 4.1 experiment with node (in black) and link IDs (in blue).

For the network scenario with correlated link costs, link 3 and link 5 are simulated to have positive correlation of 0.35. The simulated average link costs across the population of 500 agents are $\bar{c} = (1.72, 2.03, 0.71, 1.48, 1.10)$, while the standard deviations are $\sigma_c = (0.40, 0.41, 0.23, 0.42, 0.31)$. The average path costs in this correlated network are 3.20, 3.13, and 3.53, corresponding to paths 1, 2, and 3. The data are available on the GitHub site noted in Section 4 as *Test Set 2*. Based on the simulated perceived costs and same assumption as above, 43% choose (1,4), 48% choose (2,5), and 9% choose (1,3,5).

For the proposed method, an initial common prior of $c_0 = 0.5$ is assumed for all links. Algorithm 1 is employed to obtain an estimate of link costs for each of the 500 agents such that their posterior mean values are within a tolerance of 0.001 of the prior values. Since there are three routes, there are only two degrees of freedom for link costs to vary, so we do not expect estimated distributions to reflect more than two alternative options.

For the aggregate multinomial and mixed logit models, the utility functions are based on route costs to be consistent with the route choices. This is by design to contrast the outcomes of the proposed method. For the logit models, the average path costs are assumed to be known as the explanatory path cost variable $X_j$ for each alternative $j$. $U_j = \beta_j X_j + \varepsilon_j$ is an aggregate utility function that is dependent only on the same average path cost variables for everyone. $X_2$ is set to be the utility of path 2 (2,5) relative to path 1 (1,4): $X_2 = c_2 + c_5 - c_1 - c_4$, while $X_3$ is the utility of path 3 (1,3,5) relative to path 1: $X_3 = c_3 + c_5 - c_4$. In the mixed logit, $\beta_j$ is normally distributed.

*4.1.3 Results: Heterogeneity?*

We first estimate the parameters for the independent and correlated networks using the proposed method. Algorithm 1 is employed with the convergence shown in Fig. 5 for the independent network. Based on a tolerance of 0.001, the algorithm terminated after 22 iterations for the independent network data set and 19 iterations for the correlated network data set.

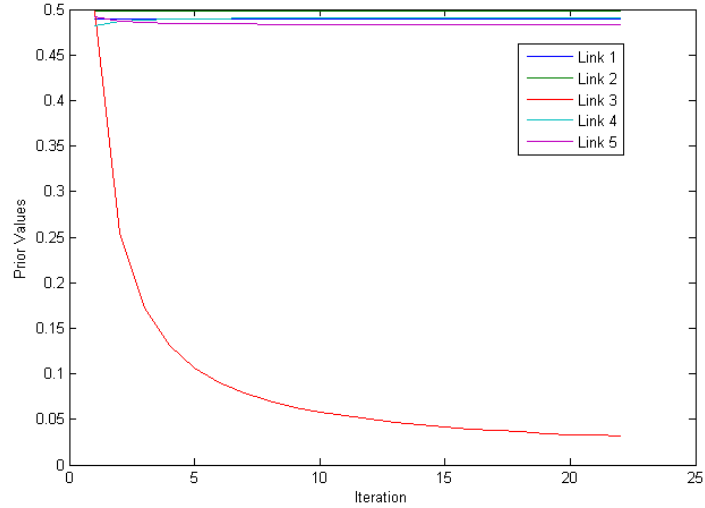

**Fig. 5. Convergence of algorithm 1 on test network for the independent link costs scenario.**

The results confirm our hypothesis. Fig. 6 illustrates how the multi-agent inverse optimization outputs a distribution of link costs across the population based on observation of their route choices and the reliance on the normative route choice behavior in the inverse transportation problem. The values are $\{0.489, 0.498, 0.009, (0.490, 0.493), (0.481, 0.484)\}$ on the independent network corresponding to links 1 to 5, and $\{(1.689,1.693), (2.061,2.065), 0.371, 1.476, 1.104\}$ on the correlated network. The link costs end up being homogeneous for the first three links in the independent network (latter three links in the correlated network), and are split over two different values for the remaining two links (the first two links in the correlated network). This reflects how even a network with only two degrees of freedom in information can lead to an estimation of heterogeneous link costs.

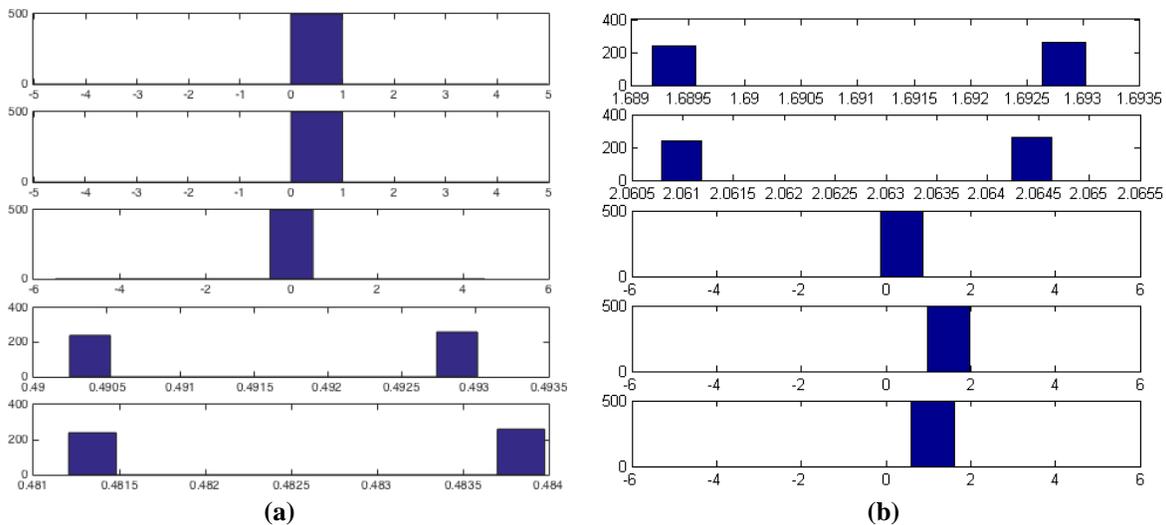

(a)            (b)

**Fig. 6. Output distribution of posterior link costs across the population of 500 simulated agents for link 1 (top) to link 5 (bottom), for (a) independent link network and (b) correlated link network.**

In summary, the tests in these scenarios verify that the multi-agent inverse transportation problems can estimate heterogeneous link costs based only on observed route choices regardless of whether the links are independent or correlated. By comparison, the study by Hong et al. (2017) looks at route level attributes only, and Chow and Recker (2012) stick to same-level link observations and link costs in activity routing.

For context, the route choices are modeled using multinomial and mixed logit models in *R* using the average route costs as the explanatory variables. The estimated multinomial logit models have log-likelihood values of $-417.24$ for the independent network and $-469.47$ for the correlated network. The McFadden R^2 values are 0.240 and 0.145, respectively. For the mixed logit model, a sampling of 100 simulated draws is used to obtain the results. Using the Broyden-Fletcher-Goldfarb-Shanno (BFGS) method, the algorithm converges to an estimate after 4 iterations. For the mixed logit model, $LL = -417.23$ and $\rho^2 = 0.240$ for the network with independent link costs. For the correlated network, the LL=$-469.47$ and $\rho^2 = 0.145$. The estimated coefficients are shown in Table 3.

Since the two networks do not have the same link cost distributions, a direct comparison of the results is not expected. However, the results clarify the value of the multi-agent inverse transportation problems when interpreted alongside one another.

- While the proposed method endogenously obtained the average link costs, the statistical models required prior information about the average path costs in order to be estimated.
- The statistical models clearly do not provide estimates of link-level parameters, much less link-level heterogeneity.
- The estimated results suggest that the standard deviations of the mixed logit models (and hence the distribution assumption for taste variation in path costs) for both the independent and correlated networks are statistically insignificant (t-stats of 0.0081 and 0.0029, respectively). Despite there being path level variation in perceived costs, it is difficult to capture this heterogeneity using the mixed logit model for this example.
- The proposed method correctly fits each individual's route choice to obtain 100% fit to the data of 48% for path 1, 48% for path 2, and 4% for path 3. On the other hand, the estimated shares from MNL are 50.2% for path 1, 45.7% for path 2, and 4.1% for path 3. Similarly, in the correlated network, the observed shares are 43% for path 1, 47.8% for path 2, and 9.2% for path 3, while estimated shares from MNL are 38.8%, 51.3%, and 9.9%.

**Table 3. Estimated parameters and significance tests for multinomial and mixed multinomial logit model**

*Multinomial Logit, independent network*

| Variable | Estimate | Standard error | t-statistic |
|---|---|---|---|
| $X$ | $-4.93040$ | 0.44997 | $-10.957$*** |

*Mixed Logit, independent network*

| Variable | Estimate | Standard error | t-statistic |
|---|---|---|---|
| $X$ | $-4.93848$ | 2.08812 | $-2.3650$* |
| $sd.X$ | 0.18917 | 23.22305 | 0.0081 |

*Multinomial Logit, correlated link network*

| Variable | Estimate | Standard error | t-statistic |
|---|---|---|---|
| $X$ | $-4.05413$ | 0.38268 | $-10.594$*** |

*Mixed Logit, correlated link network*

| Variable | Estimate | Standard error | t-statistic |
|---|---|---|---|
| $X$ | $-4.054811$ | 0.656032 | $-6.1808$* |
| $sd.X$ | 0.070375 | 24.292289 | 0.0029 |

Signif. codes: 0 '***' 0.001 '**' 0.01 '*' 0.05 '.' 0.1 ' ' 1

*4.1.4 Results: what happens when a link breaks down and the network changes?*

To illustrate the method's ability to evaluate significant structural changes in the network, we consider a scenario where one of the links fail. Scenarios 7 to 10 deal with closing link 3 for both the independent and correlated networks. Under the new scenarios, the estimated models are applied to validate their accuracy in terms of total route shares. When link 3 is closed, the alternative path 3 no longer exists, and there are only two routes choices left. Under these scenarios, the simulated observed routes show that 50% of the travelers take path 1 in the independent network, while 47.6% take path 1 in the correlated network.

The shortest path assignment using the link costs estimated with the multi-agent inverse optimization indicate with 100% fit the optimality of the observed choices. For context, the statistical models show some error as reported in Table 4. Since the mixed logit estimation was a poor fit with statistically insignificant standard deviations, that model is not applied in these scenarios.

**Table 4. Estimated shares (MNL) vs. actual shares of route choices when link 3 is closed (scenarios 7 – 10)**
*Multinomial Logit, independent network*

| Alternatives | Estimated Shares | Actual Shares | Error |
|---|---|---|---|
| Path 1 | 0.524 | 0.5 | 0.024 |
| Path 2 | 0.476 | 0.5 | 0.024 |

*Multinomial Logit, correlated link network*

| Alternatives | Estimated Shares | Actual Shares | Error |
|---|---|---|---|
| Path 1 | 0.431 | 0.476 | 0.045 |
| Path 2 | 0.569 | 0.524 | 0.045 |

*4.2 Link capacity dual price estimation parameter recovery test*
*4.2.1 Experiment 2 design and data inputs*

In the second and third experiments, the Nguyen-Dupuis (1984) network shown in Fig. 7 is used. In the second experiment, the goal is to conduct a parameter recovery test. Based on Proposition 1, the dual prices are unique and homogeneous across the population of agents. It should therefore be possible to assume link capacities on the network, solve a multicommodity flow problem to simulate the "observed" flows, and then apply Algorithm 2 to recover the dual prices.

The standard demand and link cost parameters from the Nguyen-Dupuis network is assumed: 400 travelers for OD (1,2), 600 travelers for OD (4,2), 800 travelers for OD (1,3), and 200 travelers for OD (4,3). By design, the paths in the Nguyen-Dupuis network can be easily enumerated. These are sorted by length and shown in Table 5 with the corresponding path IDs. Initial capacities of 400 at link 1 and 800 at link 7 are assumed to simulate the observed flows.

For simulating the path sampling, each of the paths is randomly drawn with probability equal to the percent flow on that path from the solution to the multicommodity flow problem. A summary of 100 sampled paths is provided in Fig. 8, and the data set is fully accessible on the GitHub site as *Test Set 3*. Although the multicommodity flow problem may require an integer solution, in this case an LP-relaxed solution is obtained revealing dual prices of $w_1^* = 7$ and $w_7^* = 5$.
The solution of the flow assignment under the hidden link capacities is used to represent the simulated observation, as shown in Fig. 1. The use of paths is dictated by ascending order of costs. For example, if an agent for OD (1,2) chooses to take path 2, it is because the dual price of path 1 has an effective value of 4 or more.

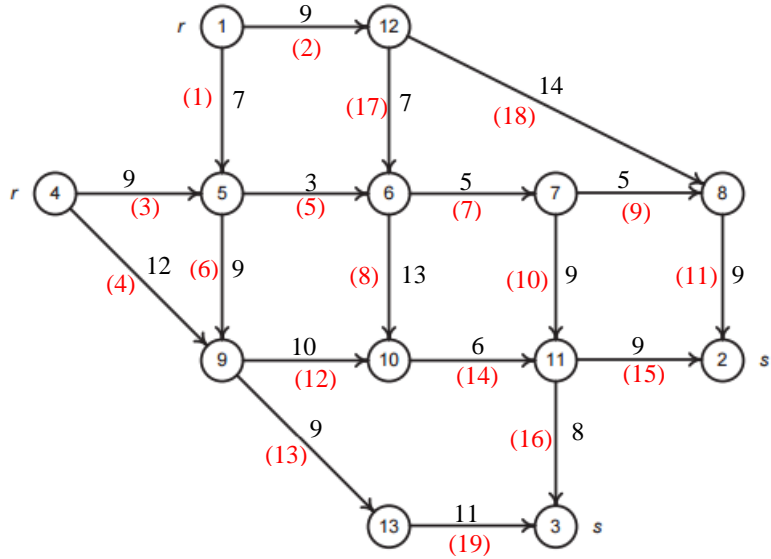

**Fig. 7. Nguyen-Dupuis (1984) network.**

**Table 5. Enumerated paths for each of the four OD pairs, sorted by length in ascending order**

| OD Pair | Path | Node sequence | Link sequence | Length | OD Pair | Path | Node sequence | Link sequence | Length |
|---|---|---|---|---|---|---|---|---|---|
| (1,2) | 1 | 1-5-6-7-8-2 | (1)-(5)-(7)-(9)-(11) | 29 | (1,3) | 14 | 1-5-6-7-11-3 | (1)-(5)-(7)-(10)-(16) | 32 |
|  | 2 | 1-5-6-7-11-2 | (1)-(5)-(7)-(10)-(15) | 33 |  | 15 | 1-5-6-10-11-3 | (1)-(5)-(8)-(14)-(16) | 37 |
|  | 3 | 1-5-6-10-11-2 | (1)-(5)-(8)-(14)-(15) | 38 |  | 16 | 1-5-9-10-11-3 | (1)-(6)-(12)-(14)-(16) | 40 |
|  | 4 | 1-5-9-10-11-2 | (1)-(6)-(12)-(14)-(15) | 41 |  | 17 | 1-5-9-13-3 | (1)-(6)-(13)-(19) | 36 |
|  | 5 | 1-12-6-7-8-2 | (2)-(17)-(7)-(9)-(11) | 35 |  | 18 | 1-12-6-7-11-3 | (2)-(17)-(7)-(10)-(16) | 38 |
|  | 6 | 1-12-6-7-11-2 | (2)-(17)-(7)-(10)-(15) | 39 |  | 19 | 1-12-6-10-11-3 | (2)-(17)-(8)-(14)-(16) | 43 |
|  | 7 | 1-12-6-10-11-2 | (2)-(17)-(8)-(14)-(15) | 44 |  |  |  |  |  |
|  | 8 | 1-12-8-2 | (2)-(18)-(11) | 32 |  |  |  |  |  |
| (4,2) | 9 | 4-5-6-7-8-2 | (3)-(5)-(7)-(9)-(11) | 31 | (4,3) | 20 | 4-5-6-7-11-3 | (3)-(5)-(7)-(10)-(16) | 34 |
|  | 10 | 4-5-6-7-11-2 | (3)-(5)-(7)-(10)-(15) | 35 |  | 21 | 4-5-6-10-11-3 | (3)-(5)-(8)-(14)-(16) | 39 |
|  | 11 | 4-5-6-10-11-2 | (3)-(5)-(8)-(14)-(15) | 40 |  | 22 | 4-5-9-10-11-3 | (3)-(6)-(12)-(14)-(16) | 42 |
|  | 12 | 4-5-9-10-11-2 | (3)-(6)-(12)-(14)-(15) | 43 |  | 23 | 4-5-9-13-3 | (3)-(6)-(13)-(19) | 38 |
|  | 13 | 4-9-10-11-2 | (4)-(12)-(14)-(15) | 37 |  | 24 | 4-9-10-11-3 | (4)-(12)-(14)-(16) | 36 |
|  |  |  |  |  |  | 25 | 4-9-13-3 | (4)-(13)-(19) | 32 |

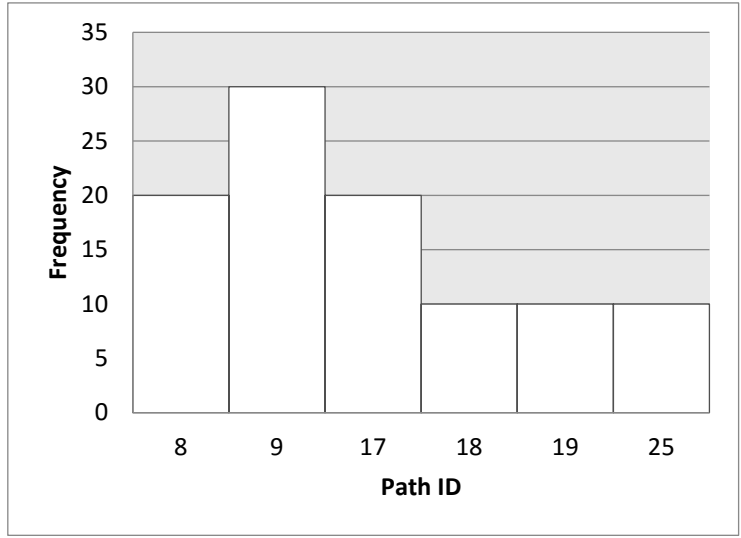

**Fig. 8. Histogram of simulated route observations in Experiment 2.**

Based on this observation, we assume there are six distinct agent groups, where all members of the group are homogeneous since no additional information is available in this experiment. Assuming that we know there are capacities at link 1 and link 7 but their values are unknown, the inverse shortest path problem $\phi^{-1}(g_1, w_0^1, x_1^*)$ is illustrated below for an agent (path 8 in Table 5) going from node 1 to node 2 from a prior of $w_0^1 = [0,0]$. The $y$ values are the dual variables of the original shortest path problem. Because the shortest path constraints are equality constraints, the dual prices here are unbounded.

$$\min \phi^{-1} = e_{1,1}^n + f_{1,1}^n + e_{1,7}^n + f_{1,7}^n$$
$$s.t.$$
$$-y_1 + y_5 \leq 7 + w_{0,1}^n - e_{1,1}^n + f_{1,1}^n$$
$$-y_1 + y_{12} \leq 9$$
$$-y_4 + y_5 \leq 9$$
$$-y_4 + y_9 \leq 12$$
$$-y_5 + y_6 \leq 3$$
$$-y_5 + y_9 \leq 9$$
$$-y_6 + y_7 \leq 5 + w_{0,7}^n - e_{1,7}^n + f_{1,7}^n$$
$$-y_6 + y_{10} \leq 13$$
$$-y_7 + y_8 \leq 5$$
$$-y_7 + y_{11} \leq 9$$
$$+y_2 - y_8 \leq 9$$
$$-y_9 + y_{10} \leq 10$$
$$-y_9 + y_{13} \leq 9$$
$$-y_{10} + y_{11} \leq 6$$
$$+y_2 - y_{11} \leq 9$$
$$y_3 - y_{11} \leq 8$$
$$y_6 - y_{12} \leq 7$$
$$y_8 - y_{12} \leq 14$$
$$y_3 - y_{13} \leq 11$$

$$e_{1,1}^n - f_{1,1}^n \leq w_{0,1}^n$$
$$e_{1,7}^n - f_{1,7}^n \leq w_{0,7}^n$$
$$-400y_1 + 400y_2 = 400(9 + 9 + 14)$$
$$e, f \geq 0$$

*4.2.2 Results*

We run Algorithm 2 to seek the corresponding dual prices that led to this flow observation. The convergence of the link capacity dual prices is shown in Fig. 9.

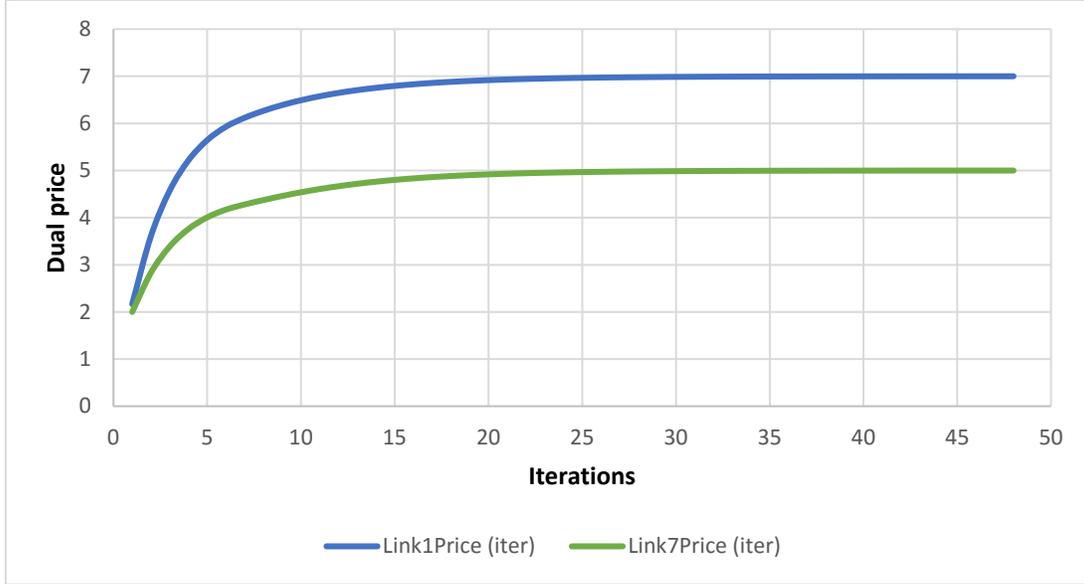

**Fig. 9. Convergence of dual prices using Algorithm 2.**

This test indicates that it is indeed possible to use the proposed method to update network parameters that influence the agents' route choices. In addition, all the agents end up with the same dual price values as the priors. This test shows how our proposed method can use individual agent learning to infer the value of shared system resources.

## 4.3 Verification of method in an online multi-agent learning environment
*4.3.1 Experiment 3 design and data inputs*

In the third experiment, we wish to verify the applicability of the proposed method as a network monitoring tool. It is assumed that data is received in real time from one agent at a time. After each agent observation, an update is conducted to learn of any changes in the dual prices in the network.

The experiment is designed as follows. We change link 7 from a capacity of 800 to a capacity of 500 and once again solve the capacitated assignment problem. In this state, the dual prices are found to be $w_1^{**} = 7$ and $w_7^{**} = 6$. Next, we randomly draw observations from the two states: the first 100 sequential samples are drawn from observations under the initial 800 capacity state, followed by 100 sequential samples under the 500 capacity state representing the capacity drop in link 7, and finally another 100 sequential samples under the 800 capacity state representing a return to initial state. The data is summarized in Fig. 10 and accessible on GitHub as *Test Set 4*. The time of each arrival is assumed to be constantly distributed to be one unit of time.

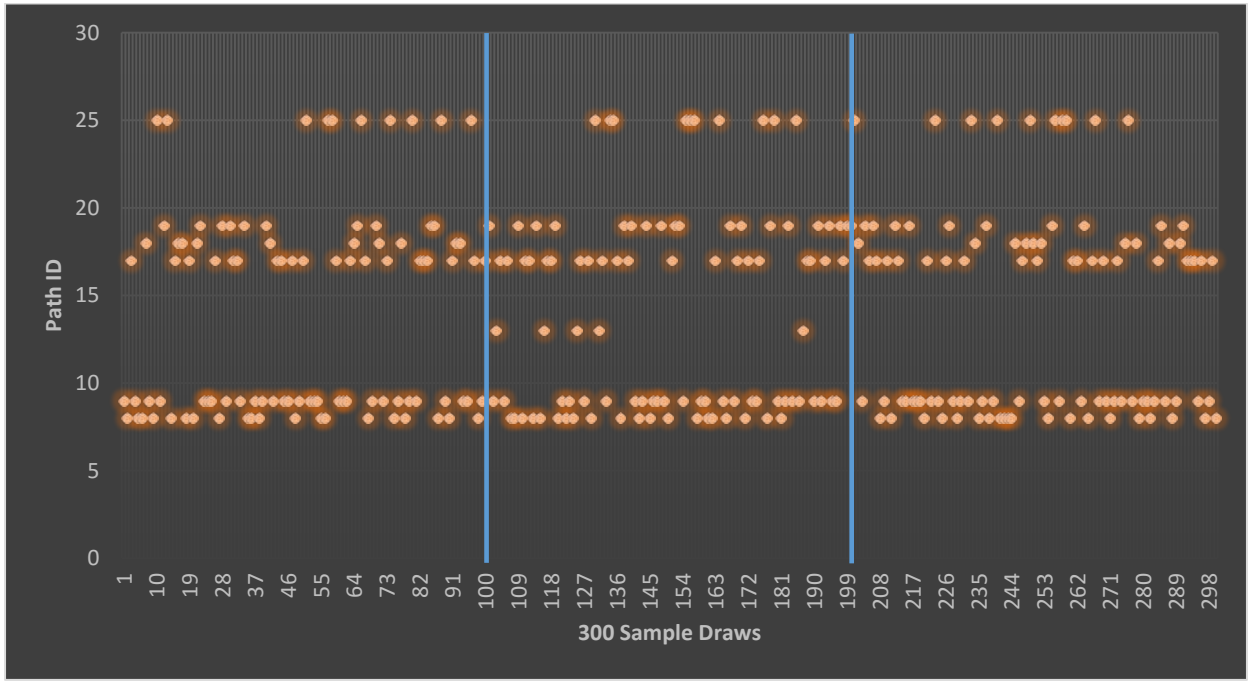

**Fig. 10. Trajectory of simulated routes observed in Experiment 3 with regime changes marked by blue vertical lines.**

Fig. 10 shows there are key paths that directly affect the dual prices when they are observed. For example, path 13 only appears when the system is operating under the 500 capacity regime, while path 18 only appears in the 800 capacity regime. We test to see if the methodology, when operated in an online learning environment, is sensitive to these regime changes.

*4.3.1 Results*

The online updating approach in Algorithm 3 is employed. Each update uses the posterior of the previous update as its prior. We get the following trajectory of the posterior dual prices shown in Fig. 11 as an example of how the monitoring occurs over 300 sequential observations.

The result shows that the proposed method is indeed sensitive to regime changes in this example, even as there is a learning period after each state change. The learning rate depends on the likelihood of the right observation that comes along to reveal the need for a change. For example, the change to 500 capacity state does not impact the monitoring of the dual price immediately. It is not until a new route observation of path 13, indicating a detour in route because of the decreased capacity, does the dual price shift. As a result, the sampling rate is important. The routes are also important. In this case, the monitoring system is able to detect a shift back and forth because the 500 capacity state leads to a different set of routes than the 800 capacity state. If the routes remain the same, no change may be detected.

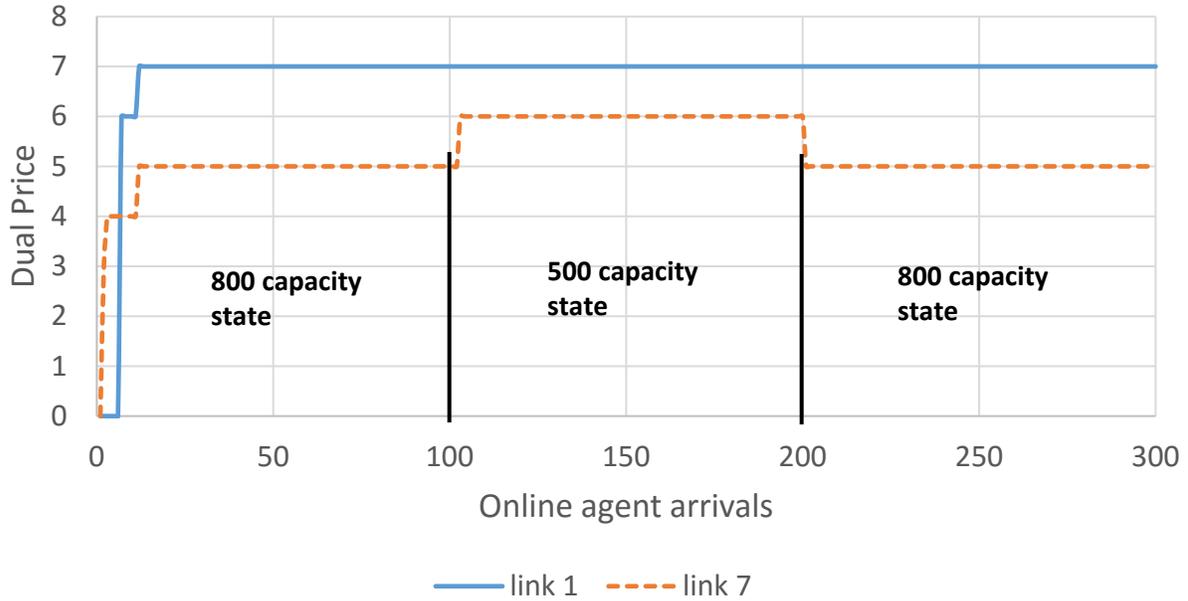

**Fig. 11. Dual price trajectories based on 300 simulated agent arrivals operating in three separate states.**

As designed, the monitoring system does not currently allow the dual prices to deflate to zero. If, for example, a link that was initially operating at capacity but is now no longer at capacity, the system would not be able to detect a lack of detour flows. One possible solution is to build in a time value component, so that sampled data will also include their inter-arrival times. Longer periods of time of inactivity would result in discounting of the dual prices back to zero. However, this would require proper calibration based on demand densities and sampling rates. We will look into this issue in future research.

### *4.4 Illustration of online network learning using real data from Queens, New York*
*4.4.1. Data and experimental design*

In this final experiment, we illustrate network learning using a real data example. A highway network from Queens in New York City is shown in Fig. 12 overlaid upon a Google Maps image. The link free flow travel times ("FF time") are presented in Table 6. The network is designed to have two entries/exits for each of the four cardinal directions. On June 5, 2017, we queried a series of shortest paths from Google Maps API based on Google's real-time travel times. The queried data, along with the network information and network learning code, are all located in the GitHub site. The following steps are taken to obtain this data.
1. Initiate with dual prices equal to zero for all links in the Queens freeway network.
2. Starting at 6:30AM, and every 5 minutes thereafter until 9:30AM,
    a. Randomly choose one cardinal direction as the origin and one as the destination.
    b. Sample the four real-time shortest paths for each of the possible entry/exit pairs. For example, if origin is North and destination is South, there are four shortest paths: N1-S1, N1-S2, N2-S1, N2-S2.
    c. Keep the one that is shortest among these as the simulated observation.
    d. Run Algorithm 3 to update the link dual prices based on the observation.

As congestion occurs in the network, the effects of the capacity on shifting routes should be recognized by the network learning algorithm. The dual prices should reflect links that become more congested with binding capacity effects that result in route diversions. The magnitudes of the dual prices should give a relative measure of the insufficient capacity in the link with respect to other links. We take snapshots of the shortest paths found in Google Maps (where congested links are typically red to black in color) so that qualitative comparisons can be made, as a comprehensive quantitative comparison is not possible with these latent variables.

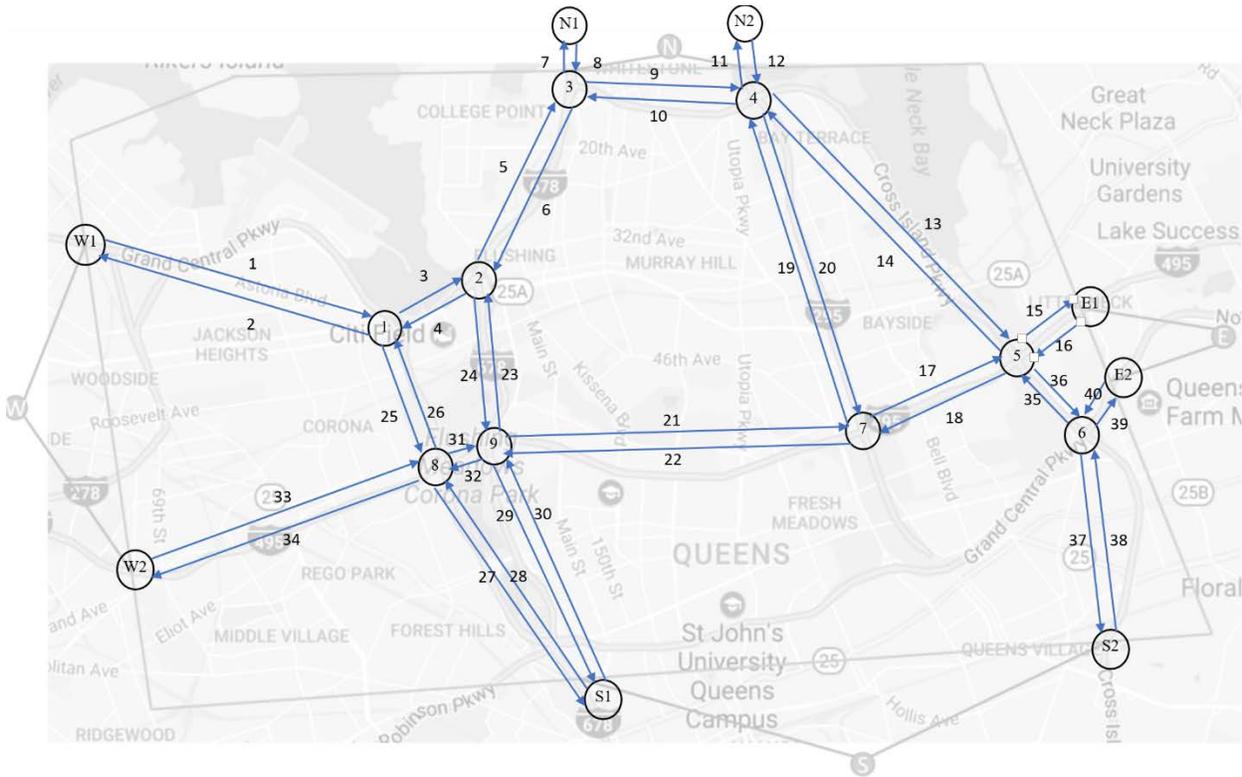

**Fig. 12. Queens freeway network.**

**Table 6. Link attributes for the Queens freeway network.**

| Link_id | start_node | end_node | FF time (s) | Link_id | start_node | end_node | FF time (s) |
|---|---|---|---|---|---|---|---|
| 1 | W1 | 1 | 211 | 21 | 9 | 7 | 233 |
| 2 | 1 | W1 | 211 | 22 | 7 | 9 | 233 |
| 3 | 1 | 2 | 77 | 23 | 9 | 2 | 99 |
| 4 | 2 | 1 | 77 | 24 | 2 | 9 | 99 |
| 5 | 2 | 3 | 133 | 25 | 1 | 8 | 95 |
| 6 | 3 | 2 | 133 | 26 | 8 | 1 | 95 |
| 7 | 3 | N1 | 39 | 27 | 8 | S1 | 180 |
| 8 | N1 | 3 | 39 | 28 | S1 | 8 | 180 |
| 9 | 3 | 4 | 113 | 29 | 9 | S1 | 180 |
| 10 | 4 | 3 | 113 | 30 | S1 | 9 | 180 |
| 11 | 4 | N2 | 50 | 31 | 8 | 9 | 36 |
| 12 | N2 | 4 | 50 | 32 | 9 | 8 | 36 |
| 13 | 4 | 5 | 228 | 33 | W2 | 8 | 178 |

| 14 | 5  | 4  | 228 | 34 | 8  | W2 | 178 |
| 15 | 5  | E1 | 54  | 35 | 6  | 5  | 60  |
| 16 | E1 | 5  | 54  | 36 | 5  | 6  | 60  |
| 17 | 7  | 5  | 109 | 37 | 6  | S2 | 101 |
| 18 | 5  | 7  | 109 | 38 | S2 | 6  | 101 |
| 19 | 7  | 4  | 206 | 39 | 6  | E2 | 57  |
| 20 | 4  | 7  | 206 | 40 | E2 | 6  | 57  |

*4.4.2. Queens freeway network experimental results*

Fig. 15 shows the trajectory of the link dual prices (the ones that became binding) as they evolve from one new sample update to the next. The figure illustrates the sensitivity of the method to changes in the network parameters over time, despite being based on only 37 randomly sampled individual route choices. We provide a snapshot of the dual prices for the links as they change every half hour throughout the 3-hour study period as shown in Fig. 14. For comparison, we include Fig. 15 to show screenshots of the Google Maps real-time shortest paths found at the same times as the dual price snapshots. The screenshots provide a qualitative validation by indicating the presence of congestion that occurs at similar segments and similar time frames.

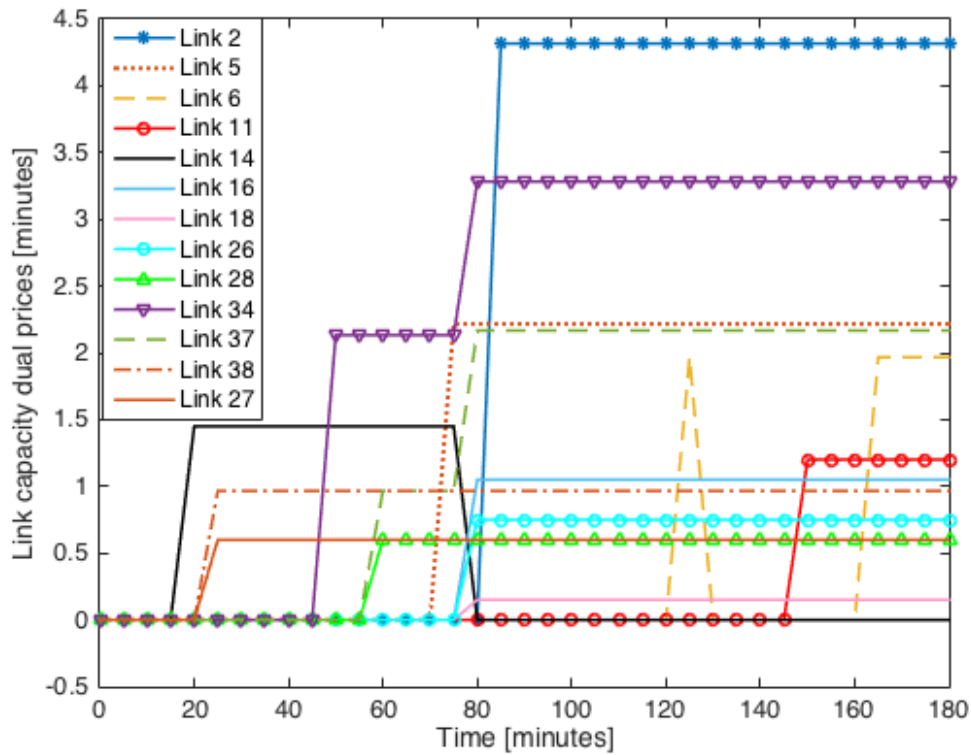

**Fig. 13. Trajectories of link dual prices as estimated using Algorithm 3 for Queens freeway network over a 3-hour period.**

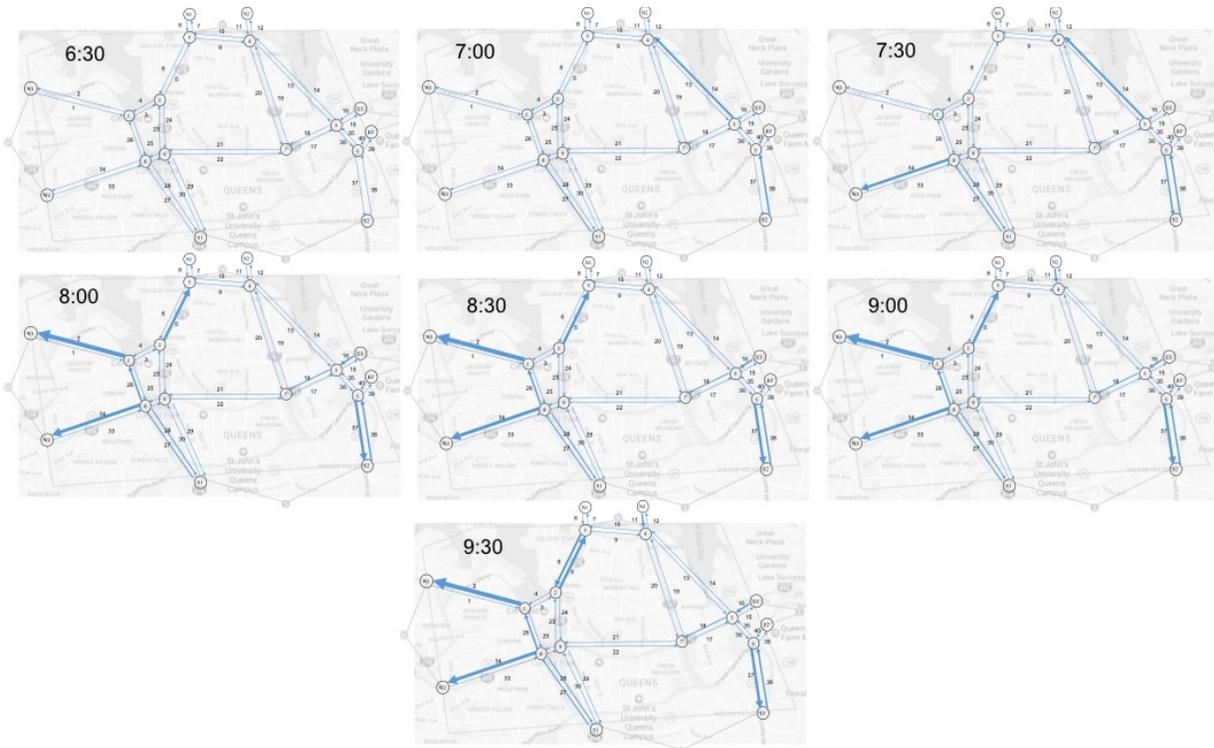

**Fig. 14. Snapshot of multi-agent IO output dual prices at every half-hour with non-zero prices represented by heavier arrows.**

A number of conclusions are drawn from this illustrative experiment.
- Network system attributes like link dual prices can be updated using only samples of individual route observations, *without need to estimate total link or path flows*. This demonstrates the significance of this methodology in being able to cheaply monitor a transportation network's system performance over time.
- The changes show that the inference model is indeed sensitive to changes in the system. As traffic increases from 6:30AM to 9:30AM in the study period resulting in more spillbacks and incidents impacting link capacities, the set of dual prices steadily increases on average as shown in Fig. 13.
- The accuracy of the inference cannot be established quantitatively. However, a visual comparison between Fig. 14 and Fig. 15 indicate similarities in positive dual prices where congestion occurs. For example, the 7:00AM screenshot shows that the segment between nodes 4 and 5 is highly congested, and that is interpreted correctly in Fig. 14. The 7:30AM screenshot reveals the alternative path traversing the link between nodes 8 and S1 is congested, which is captured correctly in the inference model. The 8:30AM screenshot indicates congestion between nodes 5 and 7, which is also captured by the model. This delay lingers through 9:30AM, and is properly captured as well by the inference model.

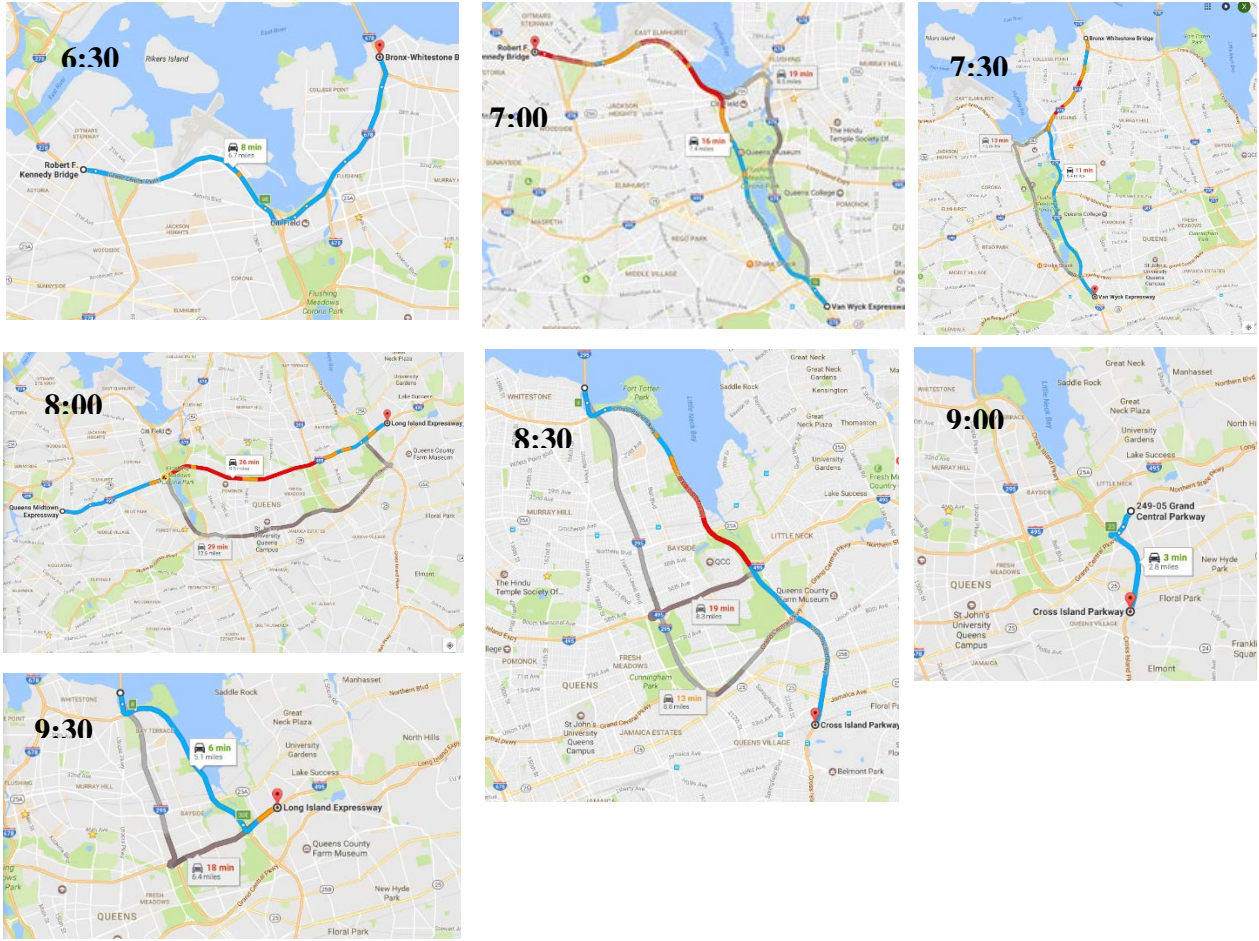

**Fig. 15.** Screenshots of Google Maps real-time shortest path input data at corresponding half-hour intervals that reveal congested segments.

## 5 Conclusion

In this study, we derived a new class of inverse optimization models such that shared network resources can be quantified from agent observations. This class is fundamentally different from the classic inverse optimization model, which requires more statistically inefficient estimation of aggregate system parameters like population link or path flows. The proposed multi-agent inverse optimization model class captures heterogeneity in agents using shared system resources, and also infers such system parameters as link capacity dual prices all without having to estimate population flows. This contribution makes it possible to cheaply apply IO techniques to many data-driven transportation problems in which observations are obtained from only samples of selfish agents or in an online learning setting.

To elaborate, we formalize a multi-agent inverse optimization modeling approach using a fixed point common prior to capture heterogeneity, relate that approach to link capacity dual prices through decomposition properties, and propose three algorithms to support these models. We further prove that the method can obtain unique dual prices for a network shared by the agent population in polynomial time (depending on LP algorithm used). The methodology is tested in four experiments:

1) For a path-enumerated 4-node network, we verify that the methodology can indeed obtain heterogeneous estimates of link cost parameters, even when there is not enough structural information for meaningful interpretation with a purely descriptive method like a mixed logit model.
2) For the Nguyen-Dupuis network, we conduct a parameter recovery test to verify that the proposed method works in inferring shared system resources through agent information. This test illustrates how much easier it is to infer impacts of link capacities than to try and estimate them directly as Güler and Hamacher (2010) tried to do as an NP-hard problem.
3) We construct an online learning example to demonstrate how the method can work in this setting. Link 7 in the network is set to experience a capacity drop before returning to its original state to mimic an incident. The results show that the online learning is able to pick up on that drop through an increase in the dual price observed from updated sequences of agent route choices.
4) The online learning is further illustrated using real data from a freeway network in Queens, New York, based on sampled real-time Google Maps shortest path queries. Through these queries, we are able to estimate the link capacity interactions with the travelers under congestion, and monitor this evolution over a three-hour period.

This work differs from other data-driven methods in the literature. Learning is made through the use of a normative behavioral mechanism so that online monitoring and strategic planning scenario evaluation are possible. The research also has implications for automated systems and artificially intelligent networks in the context of autonomous fleets (Guo et al., 2017) and smart cities. By structuring the learning in the same environment as the design and operation, it makes it easier in future research to design integrated learning and optimization strategies in networks. For example, some recent research is looking at ways to optimize resources to sense and learn from a network (e.g. Ryzhov and Powell, 2011). A next step in this evolution will be to operate a system that jointly considers resource allocation to optimally serve users and learn from them. Multi-agent IO is one way to approach such a problem.

Other extensions of this research include: conducting studies using multiple sensor sources (e.g. loop detectors, video cameras, taxi GPS data) and integrating the multi-agent IO approach; considering Bayesian techniques like Markov chain Monte Carlo methods for online learning through sampling (Tebaldi and West, 1998); and designing more sophisticated online learning systems that incorporate time value of observations and deterioration rate of dual prices. For example, a new observation that shows dual price is 5 instead of 0 can have different meanings if the observation arrives 1 minute later versus 1 hour later. This temporal component needs to be studied. Other aspects of real applications also need to be considered: data can be noisy (e.g. perceived link capacity dual price for agents may differ) and may require stochastic assignment consideration (Ashok and Ben-Akiva, 2002), only fragments of actual paths may be available (e.g. transit fare smart card data), or travelers may choose to stay at home. Aggregation methods, while discussed in Chow and Djavadian (2015), can be further expanded upon in this generalized route inference setting.

# Acknowledgments

Jia Xu and Joseph Chow were partially supported by an NSF CAREER grant no. CMMI-1652735, which is gratefully acknowledged. Preliminary results from this work were presented at TRISTAN IX in Aruba. Helpful feedback from conference attendees and three anonymous referees is much appreciated.